\documentstyle[aps,prl,twocolumn,epsf,rotate]{revtex}

\begin{document}
%% The following two lines should be there when using 'twocolumn'.
\twocolumn[\hsize\textwidth\columnwidth\hsize\csname
@twocolumnfalse\endcsname

\title{The nature of ergodicity breaking in Ising spin glasses
as revealed by correlation function spectral properties}

\draft
\author{Jairo Sinova$^{1,2}$, Geoff Canright$^{1,2}$, and A. H. MacDonald$^{2}$}
\address{$^{1}$Department of Physics,
University of Tennessee, Knoxville, Tennessee}
\address{$^{2}$Department of Physics,
Indiana University, Bloomington, Indiana}
\date{\today}
\maketitle

\begin{abstract}
In this work we address the nature of broken ergodicity in the low 
temperature phase of Ising spin glasses
by examining spectral properties of spin correlation
functions $C_{ij}\equiv\langle S_i S_j\rangle$.
We argue that more than one extensive (i.e., $O(N)$)
eigenvalue in this matrix signals replica symmetry breaking. 
Monte-Carlo simulations of the infinite-range Ising spin-glass model,
above and below the Almeida-Thouless line, support this conclusion.
Exchange Monte-Carlo simulations for the short-range model in 
four dimensions find a single extensive eigenvalue and a large 
subdominant eigenvalue consistent with droplet model expectations.
\end{abstract}


%% The following line should be there when using 'twocolumn'.
\vskip2pc]

Following Parisi's\cite{Parisi} demonstration of 
replica symmetry breaking in the low-temperature phase of 
the Sherrington-Kirkpatrick (SK) model \cite{SK}, %%G77 added back in!
there has been an ongoing debate concerning the nature of the low temperature phase
of Ising spin glasses in finite dimensions, and of glassy states in quenched-disorder
systems more generally. 
The Parisi solution breaks ergodicity through the segregation of  phase space 
into many ergodic regions, separated by energy barriers which are infinite 
in the limit $N\to\infty$
(where $N$ is the number of spins).  Each of these regions, in the case of a 
Hamiltonian invariant under global spin inversion, has an associated region related  
to it by global spin inversion, the pair forming together what we will 
refer to as a pure state pair (PSP) \cite{PSP}.
In the low-temperature phase of the SK model,
there are many PSP's, unrelated by any 
symmetry of the Hamiltonian. In accord with common usage we will
call this form of ergodicity breaking `replica symmetry breaking'
or RSB \cite{note1}.  While the Parisi solution shows every sign of being %J630
correct for the infinite-range (hence infinite-dimension) SK problem,
there is still no consensus on the form of ergodicity breaking in 
models with finite range interactions.
An alternative to the RSB scenario is 
the `droplet' picture \cite{droplet}, which postulates a single PSP in
the low temperature phase, whose excitations include rare large droplets
of overturned spins with energy (at size $L$) scaling as $L^\theta$. 

The principal numerical tool for detecting replica symmetry breaking  
has been finite size scaling of the probability
distribution function $P(q)$ for 
the overlap $q$ between the local magnetizations in different
pure states.
Several numerical studies have suggested
a behavior in finite dimensions similar to the one at the mean
field level \cite{Young83,Marinari}. Other studies using the Migdal-Kadanoff
approximation \cite{moore} and still others at zero temperature \cite{palassini}
have suggested otherwise and favor the droplet model.
None of these studies has conclusively resolved which type of broken ergodicity 
takes place in the low temperature phase in finite dimension, motivating
a search for new approaches.

\begin{figure}
\epsfxsize=3.375in
\centerline{\epsffile{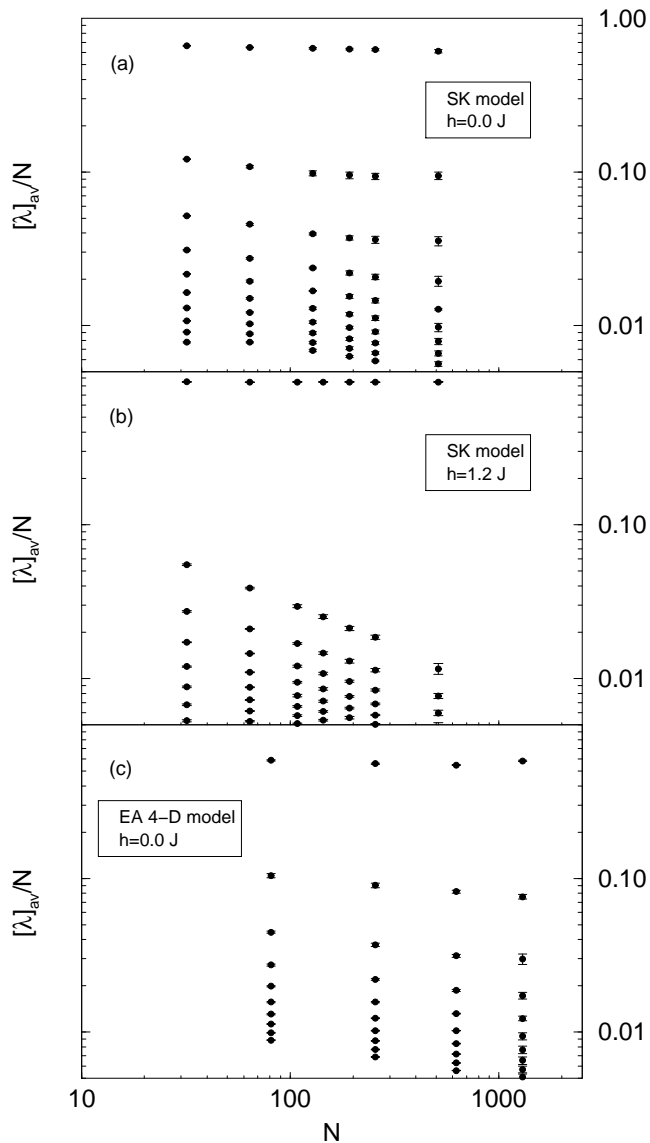}}
\caption{Scaling of the average of the ten largest
eigenvalues of $C_{ij}$  as a function of system size $N$ in (a) the SK model 
below the AT line ($h=0$, $T/J = 0.4$), (b) the SK model above the
AT line ($h/J=1.2$,  $T/J = 0.4$), and (c) the EA model in 4D
($h/J=0$, $T/J=1.0$).} 
\label{Fig1}
\end{figure}

We suggest that the nature of the low temperature phase  
can be determined by studying the spectrum of the 
spin-spin correlation function 
$\langle S_i S_j \rangle \equiv C_{ij}$.
In the case of a single PSP we will argue that this correlation
function has at most a single extensive eigenvalue.  
Thus, we take the existence of  
more than one extensive eigenvalue to be clear evidence for multiple PSPs,
ie RSB.
Our main results are shown in Fig.\ \ref{Fig1}, where we illustrate this behavior
using Exchange Monte Carlo \cite{Hukushima} simulations for the SK model, both
above (where there is a single pure state)
and below the Almeida-Thouless (AT) line \cite{AT}.
We have also used Monte-Carlo simulations to study 
Ising spin glasses in four dimensions. Here we see no signs
of RSB out to the largest size studied ($N=6^4$ spins) and instead 
find finite size behavior for the second largest eigenvalue which 
is consistent with droplet model expectations.

At a continuous symmetry-breaking phase transition, 
a suitably defined correlation function develops long-range order.
In systems without 
quenched disorder, the eigenvalues $C(\vec{k})$ of the appropriate 
correlation function
can typically be labelled by wavevector $\vec{k}$, and the long-range order is 
signalled by 
$C(\vec{k})$ becoming extensive below the transition temperature at a single value
of $\vec{k}$.
In systems with quenched disorder,
the correlation function eigenfunctions are not plane waves, but 
long-range order is still signalled by an extensive eigenvalue.
For Ising spin glasses, the correlation function matrix $C_{ij}$
is a real symmetric positive semi-definite matrix with
trace equal to the number of sites $N$,
and is analogous to the one-particle density matrix of a superfluid.  
In either system the order is off-diagonal in a position basis, and the fact
that it is long-ranged is reflected in its spectrum: an extensive
eigenvalue always signals long range order since the associated eigenfunction 
must be extended \cite{Yang}.

We now argue that there will be only one extensive eigenvalue in the case
of broken ergodicity with a single PSP.  We rewrite 
\begin{equation}
C_{ij}=\langle S_i \rangle \langle S_j \rangle 
+V_{ij}\ \ ,
\end{equation}
where $V_{ij} = \langle S_i S_j\rangle - \langle S_i \rangle \langle S_j
\rangle = \langle S_i S_j\rangle_c$ is the connected correlation function. 
If we ignore $V$, the matrix $C_{ij}$ is a pure projector $C^0 \equiv |v\rangle\langle v|$
onto the state $|v\rangle$ with $v_i = \langle S_i \rangle$. $C^0$ has one
large eigenvalue $\langle v|v\rangle = {\rm Tr} C^0 =
\sum_i \langle S_i \rangle^2$, and the remaining eigenvalues are zero. 
The influence of $V$ on the spectrum of $C_{ij}$ depends on the typical value
of its matrix elements.  If we assume that $V_{ij} \sim N^{-\delta}$ with 
$\delta > 0$, it follows from first order perturbation theory 
%(see also \cite{future}) %J630
that the largest eigenvalue will still be extensive, 
and that the largest subdominant eigenvalue $\lambda_2$ $\sim N^{1-\delta}$ at most.

In some cases we can put a bound on $\delta$.
For instance, if we simply
assume that the spin-glass susceptibility
$\chi_{SG}$ is of $O(1)$, then we get $\delta \ge 1/2$.
This should be the case above the AT line in the SK model.
The calculations shown in Fig.\ \ref{Fig1}(b) conform to this picture, with $\delta$ 
changing smoothly from $\sim 1/2$ in the figure to $\sim 1$ for larger $h$
(where the correlations approach those of a ferromagnet,
and hence $\delta \rightarrow 1$).

While the above arguments make no direct reference to spatial dimension, they
do rely on the notion of a `typical' element of $V$. This idea is certainly
appropriate for infinite-range models such as the SK problem.
We are of course most interested in finite-dimensional
problems with a spin glass phase above zero temperature.
Here the outstanding alternative to the many-valley picture 
%J630 obtained from mean-field theory  %saving space
is the droplet picture \cite{droplet}.
In this picture there are `typical' elements of $V$ which are
exponentially small, plus a set of elements of $O(1)$ in magnitude.
The fraction of the latter is of $O(1/L^\theta)$ where $L$ is the
system size and $\theta$ is a scaling exponent from the zero-temperature 
fixed point. Again we consider the `worst' case: 
supposing that $V$ for a finite sample is
dominated by one large active droplet of size of $O(L)$, then the big 
elements of $V$, appearing with probability $\sim 1/L^\theta = 1/N^{\theta/d}$,
are coherent, and so will give a large eigenvalue of order
$N\times N^{-\theta/d} = N^{1-\theta/d}$. (We have verified this
with simple numerical experiments.) If $V$ is incoherent then
its eigenvalues should grow more slowly (or decay) with $N$. Hence,
given the principal scaling assumption of the droplet picture, we expect 
the second eigenvalue $\lambda_2$ of $C_{ij}$ to grow as
$N^{1-\theta/d}$ or slower.

It follows from the above arguments that as long as there is one PSP
(or one pure state), there can be no more than one eigenvalue of $C$ which is
of order $N$. Thus
the observation of more than one such eigenvalue directly implies 
multiple PSPs. 

The SK model is a perfect candidate to test our ideas regarding
multiple large eigenvalues of $C_{ij}$, since we can tune the nature 
of the broken ergodicity by simply varying the external magnetic field $h$.
We have performed numerical calculations using Exchange Monte 
Carlo \cite{Hukushima}, which allows for faster jumps across large
energy barriers.
Our criteria for equilibration
follow closely those used in Ref. \cite{Marinari}. We check
for agreement between two different ways of calculating $\chi_{SG}$; 
one method  \cite{Bhatt&Young87}  uses the averaging of the overlap of 
two uncoupled replicas, and the other uses the standard 
thermal averaging in Monte Carlo simulations.  We  also  
check the symmetry of $P(q)$ and the variance of the eigenvalues of $C_{ij}$.

\begin{figure}
\epsfxsize=3.3in
\centerline{\epsffile{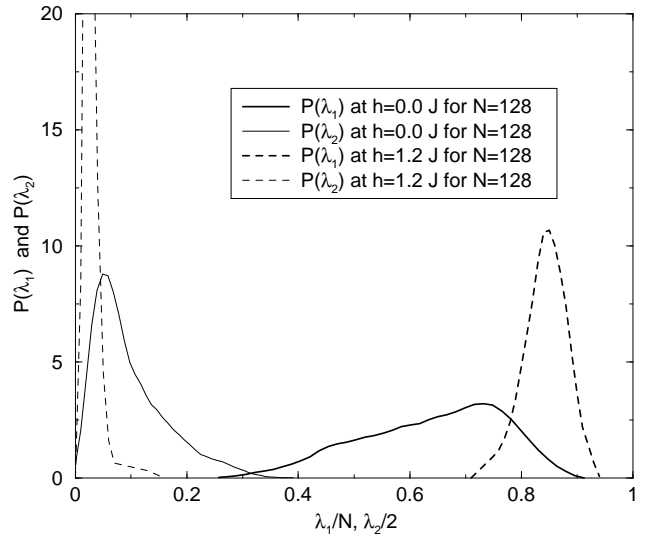}}
\caption{Distribution of the first (thick lines) and second (thin lines)
eigenvalue of $C_{ij}$  in the SK model at 
$h/J=0$ (solid line) and $h/J=1.2$ (dashed line), 
above the AT line. Here $T/J=0.4$ and N=128.}
\label{distrib}
\end{figure}

We find that the probability distributions of the eigenvalues of $C$
are broad (at least for the first two eigenvalues) below
the AT line, and very sharp above it. We illustrate
this in Fig.\ \ref{distrib}.
The breadth of these distributions indicates the need to do
disorder averaging over a large ensemble (300-2000), at least 
in the spin glass phase.
In Fig.\ \ref{Fig1} we show the scaling of the disorder average 
of the ten largest eigenvalues as a function of system size, both below 
[Fig.\ \ref{Fig1}(a)] and above [Fig.\ \ref{Fig1}(b)] the AT line
in the SK model.
It is clear from Fig.\ \ref{Fig1}(a) that at least 
two eigenvalues are of $O(N)$ for $N \agt 100$. 
We expect\cite{caveat} further $O(N)$ eigenvalues to emerge for larger
$N$, as suggested by the behavior of $[ \lambda_3 ]_{av}$ in the figure.
In contrast, there is only one large eigenvalue in Fig.\ \ref{Fig1}(b).
We find further that, above the AT line,
$[ \lambda_2]_{av}/N \sim N^{-0.52}$, consistent with
our previous arguments.
MC results for larger $h$ show that
$[\lambda_2]_{av}/N$ decays with a larger exponent (we have observed up to
$\sim 0.75$), which we expect to approach 1 for large enough $h$.

\begin{figure}
\epsfxsize=3.375in
\centerline{\epsffile{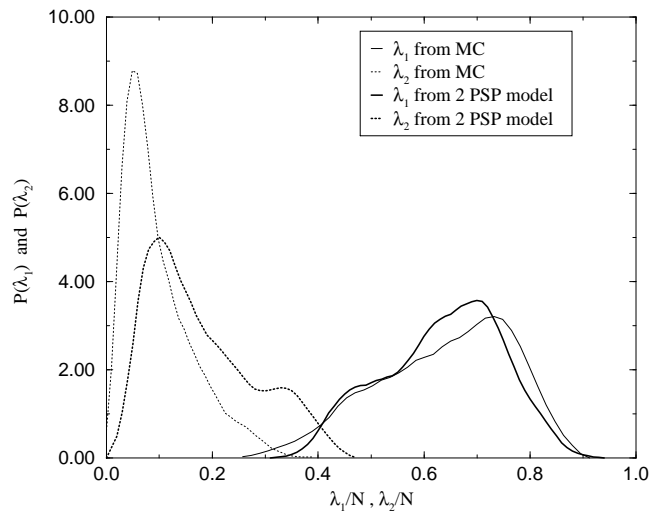}}
\caption{Distribution of the first (solid line) and second (dashed line)
eigenvalue of $C_{ij}$ obtained from the MC simulation of the SK model at
$N=64$ and $T/J=0.4$ (thin line), and the respective distributions
obtained from the two pure state model simulation (thick line) with
$\alpha = 1/2$.}
\label{nicefit}
\end{figure}

Fig.\ \ref{Fig1}(a) suggests that the 
spectrum of $C_{ij}$ for the SK problem is dominated by two
large eigenvalues (for the sizes considered here). 
We can reproduce this behaviour with the following simple model.
Suppose that phase space consists of only two spin
configurations 1 and 2, and ignore all others.
Take $C(\alpha)= \alpha C^{(1)}+(1-\alpha)C^{(2)}$, with $C^{(1)}$ and $C^{(2)}$ 
corresponding to the $C_{ij}$ of the two different configurations 
$S^{(1)}_{i}$ and $S_{i}^{(2)}$ at zero $T$, and $\alpha$ (a thermodynamic
weight) ranging from 0 to 1/2. The overlap between the two states 
is given by $q_{12}=\sum_i S^{(1)}_{i} S_{i}^{(2)}/N$.
It can be easily shown that this matrix has only two
nonzero eigenvalues, corresponding to  
\begin{equation}
\frac{\lambda_\pm(q_{12},\alpha)}{N}
=\frac{1\pm \sqrt{q_{12}^2+(1-q_{12}^2)(2\alpha-1)^2}}{2}\,\,.
\label{lambdapm}
\end{equation}
Note that $\lambda_{\pm}/N$ range from $1$ and $0$ at $\alpha=0$ to 
$(1 \pm |q_{12}|)/2$ at $\alpha=1/2$. 
It is clear from (\ref{lambdapm}) that the two largest eigenvalues
can vary over a wide range, while still in general
remaining of $O(N)$ \cite{note}.
Hence (as we saw numerically) it is important
to study the probability distribution for each eigenvalue $\tilde{P}(\lambda_i)$, 
rather than the distribution of the
eigenvalues for a single disorder realization. 
At this low level of approximation we already see
that the probability distributions of the first two eigenvalues will be
very broad in the frozen phase whenever the probability distribution
of $q_{12}$, $P_{12}(q)$, is broad. 
This simple picture however lacks 
any finite temperature and correlation effects. To introduce temperature
in the model we let
$S_i^{(1)}$ and $S_i^{(2)}$ become gaussian random variables.
We decompose $P(q)$ from our MC data as $(1/2)(P_{11}(q) + P_{12}(q))$,
with $P_{11}(q)$ (the self-overlap) a Gaussian which determines the mean
and variance of $S_i^{(1)}$ and $S_i^{(2)}$.
We then adjust their relative distribution to 
match $P_{12}(q)$. This gives us a joint distribution for $S_i^{(1)}$ and $S_i^{(2)}$
which we use to generate a distribution of matrices $C(\alpha=1/2)$.
The eigenvalues may then be obtained either by an analytical
perturbation approach 
or by direct numerical
diagonalization of the matrix $C$.
Fig.\ \ref{nicefit} compares the resulting eigenvalue distributions 
for $N=64$, with those obtained directly from the MC 
results for the SK model.
It is striking how closely they resemble one another 
in qualitative and quantitative behavior, given the simplicity of our two-state model.  

Thus we find a clear confirmation of our ideas in the behavior of the SK model, for which
we know the nature of the ergodicity breaking in the various equilibrium phases.
We now use these ideas to study
the four-dimensional Ising spin glass or Edward-Anderson (EA) model,
with nearest-neighbor interactions and a Gaussian distribution of the $J_{ij}$'s
with variance $J$.
In Figure \ref{Fig1}(c) we show the scaling of 
$[ \lambda_i ]_{av}/N$ for the first ten eigenvalues of $C$ at 
$T/J=1.0$ and $h/J=0$. 
These runs are rather deep into the frozen phase, 
since\cite{RBY90} $T_c \approx 1.75 J$.
We choose this low temperature in order to 
work as far as possible from the critical region.
This low temperature, plus the time involved in measuring $C_{ij}$,
has limited us thus far to $L \le 6$.
(We note that, even at this low temperature, 
we are still not fully out of the critical region, according to
Ref.\ \onlinecite{Bokil}.)
In this region of size and temperature we find, as in the SK spin glass phase,
a broad and asymmetric distribution of $\lambda_1$ and $\lambda_2$,
similar to that for the frozen phase illustrated in Fig.\ 2. 

We have studied both $[\lambda_i]_{av}$ and $(\lambda_i)_{typ}$ for
these distributions (the latter defined as $\exp([\ln \lambda_i]_{av})$).
Fig.\ 1(c) shows $[\lambda_2]_{av}/N$ decaying with $N$. 
On a double-log scale we see a clear straight line with exponent
$\sim 0.11$. The behavior for $(\lambda_2)_{typ}$ is similar,
except the exponent is larger, $\sim 0.15$.
This latter exponent is consistent with our lower bound for
$\delta$, given that estimates of $\theta$ in 4D range 
\cite{Hukushima,Hartmann} from 0.6 to 0.8.
Hence these data appear to fit the droplet scenario,
with a single PSP in the frozen phase. 
We cannot of course rule out the possibility that the
$[\lambda_2]_{av}/N$ points (and others for $i>2$) may flatten out 
at larger $N$, as occurred for the SK case at about $N=100$. 
Only further work at larger $N$ can help with this question. 
Also, while it is consistent that $(\lambda_2)_{typ}/N$ decays
as expected according to our arguments and the droplet picture,
the slower decay of $[\lambda_2]_{av}/N$---which clearly results from
the skew in the distribution---needs some understanding.

In summary, we have proposed, and numerically tested, a 
method for identifying the nature of ergodicity breaking in 
Ising spin glasses.  We believe \cite{caveat,note}
that RSB occurs if and only if 
the spin correlation function has multiple large
eigenvalues.  Our MC results for the infinite-range SK
problem show a clear qualitative difference in the long-range order
above and below the AT line. Because of the generality of these ideas, we
expect that they may be applied to other 
problems in statistical
physics for which RSB is a possibility.
We have also presented results for the EA model in four
space dimensions. Because of the finite-size effects plaguing all
spin-glass simulations, these results cannot
be viewed as conclusive.  However our results over the entire range 
for which we can reliably compute both thermal and disorder averages tend to
support the droplet picture, with one pure-state pair: we see no sign of multiple
large eigenvalues, and $(\lambda_2)_{typ}/N$ decays roughly in line with
droplet model expectations.  Future work based on this approach should
shed significant further light on the nature of ergodicity breaking in 
finite-dimensional spin glasses.

The authors acknowledge helpful discussions with H. Castillo,
J. Hu and E. Sorensen. This work was supported by the National Science
Foundation under grants DMR-9820816 and DMR-9714055.

\end{document}